# A METAMODEL FOR DESCRIBING THE OUTCOMES OF THE MANA CELLULAR AUTOMATON COMBAT MODEL BASED ON LAUREN'S ATTRITION EQUATION

M K Lauren

February 2005

# A METAMODEL FOR DESCRIBING THE OUTCOMES OF THE MANA CELLULAR AUTOMATON COMBAT MODEL BASED ON LAUREN'S ATTRITION EQUATION

M K Lauren

February 2005

Abstract:

In 1999 the author suggested an equation to describe the rate and temporal distribution of battle casualties. This equation was proposed as a replacement for the Lanchester equation. Unlike Lanchester's equation, the author's equation is based on a measurement of the spatial distribution of forces involved in a battle, and incorporates this by use of the concept of the fractal dimension. In this report, the form of the expected Loss-Exchange-Ratio function based on this equation is determined. It is shown how this function can be used as a Metamodel to describe the outcome of some cellular automaton models. While this is not an exhaustive proof of the validity of the equation, the match for the case studied is convincing, pointing to the potential of the equation to be adopted more generally for combat models. This is particularly so in the context of earlier work studying historical battle data that also supports an equation of the form of the author's proposal.

# EXECUTIVE SUMMARY

## Background

This study is a follow-on to previous investigations into the theory of combat modelling using fractal analysis. This body of work has received significant interest from overseas experts in the field, particularly in the US, UK and Australia. There are now many examples of this work being cited by our collaboration partners, with one notable example being a new book on complexity science and network-centric warfare by a leading UK theorist. It is envisaged that this paper and a planned future paper will tie these ideas together in such a way that it is possible that the presented theoretical framework could be used to replace many ideas in existing aggregated combat models.

## Sponsor

NZDF HQ, under core DTA Research Project D0202.

## Aim

To demonstrate the validity and usefulness of the author's 1999 equation, and to demonstrate its ability to be used as a Metamodel for the MANA cellular automaton combat model. It is hoped that this work will make a significant contribution to the scientific community and literature in the area of aggregated models of combat.

## Results

It was shown how the equation proposed by the author could be used as a Metamodel for the expected Loss-Exchange-Ratio (LER) for at least some cases of the MANA cellular automaton combat model.

What was particularly interesting for the scenario examined in detail in this report (the "Meet" scenario) was that the experimental MANA runs revealed that the LER maintained a value of around 1 for several different force ratios (i.e. number of Blue over number of Red).

While the Metamodel described this phenomena well, the experimental results are at odds with the Lanchester theory. According to Lanchester, the LER ratio should change as the force ratio changes, because the potential firepower of one side has changed relative to the other. However, this is not observed to happen in the MANA model experimental data, in fact, any change is in the opposite direction.

The author's model explains this in terms of the fractal dimension of each side changing as the force ratio is changed.

At the very least, the Metamodel appears to be useful for describing scenarios where at least one side has a complicated spatial distribution. This may in turn allow investigation in the value of dispersed forces, guerilla and swarming tactics.

Further investigation is needed to determine the extent of applicability of the model, and into how or whether it is appropriate to incorporate effects like area fire into the model.

**CONTENTS**

# 1 INTRODUCTION

## 1.1 Background

This report explores the validity of the equation suggested by the author (1999, 2001, 2002a, 2002b) for describing combat outcomes. This equation describes the rate and distribution of casualties for a battle. In this sense, it is a similar kind of equation to the well-known Lanchester equation[1]. However, despite a superficial similarity to the Lanchester equation, the formulation is quite different. The equation has the form:

$$\left\langle \frac{\Delta B}{\Delta t} \right\rangle \propto R\, k_R^{E(D_R)}\, \Delta t^{-F(D_R)}, \quad E + F = 1 \quad (1)$$

where $B$ and $R$ are the number of Blue and Red combatants, $k_R$ is the killing rate of a Red combatant, $E$ and $F$ are unknown functions and $D_R$ is the fractal dimension of the distribution of the Red side. Here, the angled brackets represent an ensemble average, so that the equation is statistical rather than a differential equation like the Lanchester equation.

Equation 1 hints at a neat phenomenological model of combat as a self-organising system. It may be imagined that combat fronts (the interface between two groups of combatants) become distorted during the events of the battle. The system of combatants can be viewed as neither rigidly ordered nor completely disordered during this period. The dynamics are in at least some part driven by short-range rules of interaction between individuals rather than via some top-down command process. Such dynamics are referred to as self-organising, and allow the combat system to be adaptable without completely losing order.

It is expected that, as a result of this self-organisation, the front becomes distorted in such a way that it is most appropriately described by a fractal (noting that most self-organising systems exhibit fractal patterns). Hence the degree of distortion of the front can be expressed as a fractal dimension. In such a scheme, a non-integer value for the fractal dimension indicates the degree to which the front is an "irregular" curve.

The equation is therefore intended to describe the rate of attrition during a period in which the $B$ Blue and $R$ Red combatants are constantly within close enough range of each other to self-organise as a result of the other side's presence.

While this makes a nice phenomenological model, the aim of this work is to explore methods for obtaining more quantitative predictions of combat behaviour from Equation 1. This was done by re-examining how measurable quantities within cellular automaton combat models might be directly used to estimate the likely outcome of a modelled battle.

Previous investigations by Lauren and Stephen (2002a) and other authors, particularly Moffat and Witty (Moffat and Passman, 2004; Moffat, Smith and Witty, 2004; Moffat and Witty, 2003; Witty and Moffat, 2003; Moffat, 2003), have demonstrated that some of the features one might expect to find in combat data if Equation 1 holds do indeed exist. In particular, it has been demonstrated in Lauren and Stephen (2002a) that the distribution of casualties from historical battles exhibit the power-law nature that the equation implies.

---
[1] Note that the Lanchester equation appears in Section 3.2 for those unfamiliar with its form.

Furthermore, it has been shown with cellular automata models, particularly the MANA and ISAAC models, that combatants tend to self-organise into fractal-type patterns, and that under these circumstances the rate of attrition can appear to behave as a power-law function of *k* (Witty and Moffat, 2002).

However, this previous work, while supporting the power-law nature of Equation 1, has so far failed to demonstrate the relationship between attrition rate and the power exponent $F(D_R)$, or even between $F$ and $D_R$.

Here it will be shown how the expected Loss-Exchange Ratio (LER, defined as the number of Blue casualties over the number of Red) appears to be dependent on $D_R$, at least for the cellular automaton model examined here. A formulation is produced which estimates the LER in terms of the fractal dimension of the distribution of the sides. This formulation thus acts as a Metamodel for the more complex cellular automaton model used.

**1.2 Spectra of casualty data**

Before examining how the LER function can be obtained in terms of the fractal dimension, *D*, it is worth discussing approaches one might consider to measuring the parameters in Equation 1 directly.

As noted in the introduction, previous investigations into the relationships between *E*, *F*, *D* and attrition rate have not been fruitful. Rather, it has simply been shown that analysis of historical data and model simulations are not inconsistent with Equation 1.

Here we briefly discuss these difficulties. In principle, it is possible to measure the values for *E* and *F* directly. However, measurement of parameters for Equation 1 is not as straightforward as it might be for a differential equation such as Lanchester's equation, due to its statistical nature.

One approach is to recast Equation 1 as a second-order statistical moment, so that:

$$\left\langle \left| B(t+\Delta t) - B(t) \right|^2 \right\rangle \propto k_R^{2q} \Delta t^{2r}, \quad -q+r=0 \quad (2)$$

and attempt to estimate the new value *r*, and hence *q*. To do this we can use the Wiener-Khinchine relation which states that for $1 < \beta < 3$

$$|f|^{-\beta} \leftrightarrow |\Delta t|^{\beta-1}$$

are Fourier transform pairs. Hence one would expect the frequency spectrum of a casualty time series to display a spectrum of the form:

$$E(f) = |f|^{-\beta} \qquad (3)$$

where $\beta = 1 + 2r$.

As already noted, Equation 3 has been observed to hold for both historical and simulated casualty data (Lauren and Stephen 2002a). One might then expect that it should be a simple matter to measure the value of $\beta$ and hence obtain *r* and *q*, then to compare these values with both the attrition rate and the fractal dimension of the distribution of each side.

However, in practice it proves difficult to robustly estimate $\beta$. There are two reasons. The first is that it is not straightforward to estimate $\beta$ from a plot of the spectrum. Recalling that on a log-log plot a spectrum of the form of Equation 3 will display a straight line, the slope of which is $\beta$, it is typically found for casualty data that the region of the spectrum which obeys this law can be relatively small, and may tend to display a slightly curved nature at the start and finish so that it is ambiguous as to where to measure the slope.

The second problem is that often the time series of casualties is sparsely "populated", particularly for simulations due to the small number of combatants. For example, in the simulation used later in this report, there may be a combined total of 100 casualties in a 200-time step simulation. In this case, the spectra typically either display poorly defined power-laws, or may simply resemble white noise.

While historical data have been demonstrated to display convincing power-law spectra, for that data there were many (often hundreds) of casualties per time step (where a time step is a day). The difficulty with reproducing this within a combat model is in modelling large numbers (thousands) of soldiers fighting complex battles.

A more promising alternative for generating model data is to use "contact" data, as was done in Lauren and Stephen (2002a). In this case, the time series describes the number of enemy entities seen per time step rather than casualties. This produces a richer time series, and it may be argued that the number of contacts ought to be proportional to number of casualties. However, though this method produces much more convincing power spectra, the connection between the spectral slope $\beta$ and the rate of attrition is still not apparent from the experimental data.

Therefore other methods are needed.

**1.3 Dependence of attrition on *k***

An alternative to trying to determine the value of *r* in Equation 2 by using the power-law exponent $\beta$ from the spectra of casualty time series is to examine how the attrition rate depends on the value of *k*.

As with the spectral analysis approach, this methodology has been tried in previous reports (Lauren, 1999; Lauren, 2002a; Witty and Moffat, 2002; Moffat, 2003). It involves finding the average time for one side to suffer a certain number of casualties for a given value of *k* for the other side. This average time is then plotted against *k* on a log-log graph to theoretically give a straight line, the slope of which is *E*.

While this previous research has suggested that attrition rate does approximate a power-law dependence on *k*, once again there are problems determining *q* from the slope displayed on these plots. An example is shown in Figure 1.

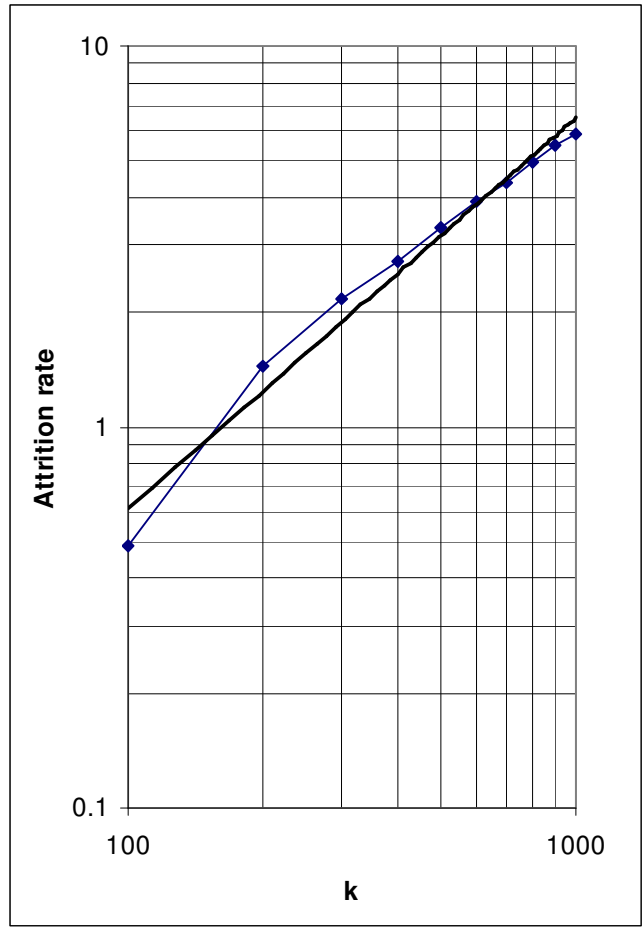

**Figure 1: Rate of attrition as a function of the opposing side's *k*. While the plot approximates the theoretically expected straight line, determining what the slope of this line should be is problematic.**

The principal difficulty is that, as with finding the power-law exponent for the spectrum, it is difficult to decide whereabouts the straight line should be fitted due to slight curvature in the data points.

Note, however, that the existence of this curvature does not necessarily show that a power-law type relationship is not valid, nor does it disprove Equation 2. As has been pointed out previously (Lauren 2002b), one would not expect the plot to exhibit a perfectly straight line. There are at least two reasons for this:

Firstly, above some value of *k*, targets are eliminated by shooters almost immediately after contact. In this case, neither force has the opportunity to manoeuvre and self-organise. Additionally, increasing *k* further has little effect on the attrition rate.

Secondly, as *k* approaches this critical value, there is less time for either side to self-organise; hence one might expect the fractal dimension *D* to have a slightly different value depending on *k*. This would suggest that the plot of attrition rate against *k* ought to curve slightly on a log-log plot, since *D* is not constant with *k*.

In summary, our attempts to measure *E* and *r* directly using these methodologies have not produced useful quantitative results to date.

**1.4 Determining the fractal dimension *D***

For this report, the fractal dimension was determined by using a box-counting technique. This involves splitting the MANA-model battlefields into four equal squares and counting the number of squares containing either Red or Blue automata, depending on the side for which the fractal dimension is being calculated. Then each of these squares is split into four, with automata-containing squares counted again, and so on. This is illustrated schematically in Lauren (2001).

The "box-counting" fractal dimension, *D*, is then given by:

$$D = \lim_{d \to 0} \frac{\log N}{\log\left(\frac{1}{d}\right)}$$

where *d* is the width of the box, and *N* the number of boxes required to cover all the automata. Note that this is equivalent to finding the slope of a straight line fitted to a plot of *N* versus *d* on a log-log plot.

Figure 2 shows an example of such a plot using data generated from the MANA model, using a spatial distribution of automata similar to that shown for the Meet scenario in the later sections. As can be seen from the figure, in this case the slope was determined by fitting a straight line to points 2 to 5. As with the discussions in the preceding sections on fitting straight lines to spectra and for attempting to directly determine the value of *E*, it is not obvious from Figure 2 which points to include in the fit. This is particularly so since there appears to be a "knee" in the plot at about point 5, so that for the points beyond this, the slope becomes substantially less steep.

Generally one would expect a "truncated" range of scales for which this power-law behaviour applies. At the smallest scales, the size of the boxes have become smaller than the characteristic distance between the automata. Hence all the automata already lie individually in boxes, so that reducing the box size further does not increase *N*. For the largest scales, a single box covers all automata, hence increasing box size further also has no effect on *N*.

However, while this difficulty in fitting a line to a log-log plot was problematic in the context of determining *q* and *r*, it was found that finding the "exact" fractal dimension was not so critical for estimating the expected LER. Rather, what was important was to be consistent about the method for estimating *D*. Thus if the estimate for $D_R$ was based on a fit of the 2nd through to the 6th points, then so too should the estimate for $D_B$.

Note that the value of *D* used in the Metamodel estimations of LER was in fact *D* averaged over all time steps. That is to say, *D* was calculated at each time step and the average value used.

In the following section, it will be shown how these estimates for *D* were able to be used in conjunction with Equation 1 to construct a Metamodel describing expected LERs for at least one MANA model scenario.

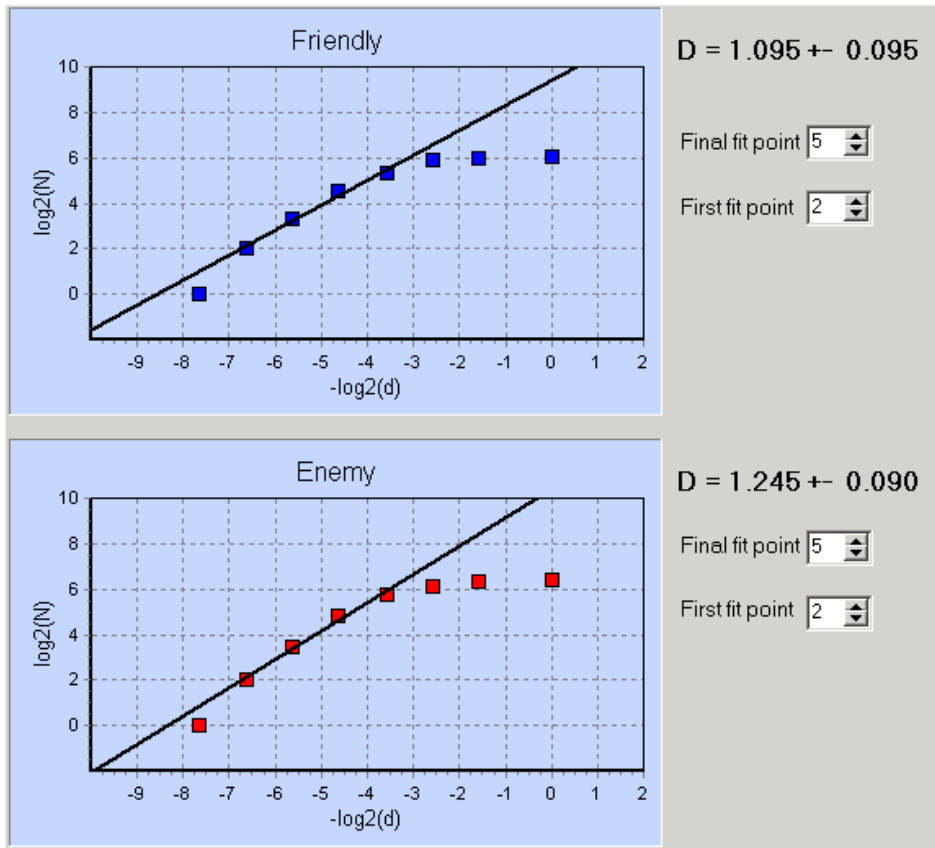

**Figure 2: Determining the fractal dimension by box-counting techniques involved finding the slope found by plotting number of boxes containing automata (*N*) against the size of the box (*d*).**

## 2 METAMODEL FOR DESCRIBING CASUALTIES IN A CELLULAR AUTOMATON MODEL

### 2.1 Calculating the LER from *D*

In this section, it is shown how the LER can be estimated given values for *D* of each side. This is done by assuming the relationship between $\beta$ and *D* as given by Moffat (Moffat, Smith and Witty, 2004)

$$\beta = D + 1.$$

Thus from Equation 2, one would expect that

$$r = D/2 \qquad (4)$$

Recalling that $-q + r = 0$, it follows that

$$q = D/2 \qquad (5)$$

In the previous sections, the emphasis was on determining the attrition rate given certain values for *E* and *F*. However, the attrition rate may not be particularly useful from an analyst's point of view, and may be skewed for various reasons. For example, at what point is the battle judged to have begun? How much of the initial manoeuvring is counted as being part of the battle? When is the battle judged to have finished?

Here, we attempt to construct a Metamodel from Equation 2 to describe the typical LER of a battle between groups of automata in the MANA model. Given that the LER is simply the difference in the size of the Blue force over the difference in size of the Red force after some time interval, one might expect that the LER is of the form:

$$\frac{\langle B(t+\Delta t) - B(t) \rangle}{\langle R(t+\Delta t) - R(t) \rangle} = \frac{R k_R^{q_R} \Delta t^{r_R}}{B k_B^{q_B} \Delta t^{r_B}} \qquad (6)$$

or, substituting Equations 4 and 5 into Equation 6:

$$\frac{\langle B(t+\Delta t) - B(t) \rangle}{\langle R(t+\Delta t) - R(t) \rangle} = \text{LER} = \frac{R k_R^{D_R/2} \Delta t^{D_R/2}}{B k_B^{D_B/2} \Delta t^{D_B/2}}. \qquad (7)$$

This equation forms our Metamodel for comparison with MANA simulations in the following sections.

### 2.2 The MANA cellular automaton model

The validity of the Metamodel was tested by comparing the outcomes of abstract combat models within the MANA (Map Aware Non-uniform Automata) cellular automaton model (Version 3), developed by the Defence Technology Agency, New Zealand (Lauren and Stephen 2002b).

Within this model, automata choose their moves according to personality weightings in a similar way to the ISAAC model described by Ilachinski (2000). The movement algorithm works by calculating the penalties associated with an automaton moving to any of the cells surrounding it (see Gill 2004). The move is chosen by randomly selecting a cell with a penalty lower than some threshold, hence automata do not necessarily make "perfect" moves each turn. Penalties are smallest for moves that bring the automata nearest to objects for which they have the greatest weighting.

The order in which automata move is randomly selected each turn, and no more than one automaton can occupy a given cell. For shooting, no systematic target selection is modelled. Instead, targets are chosen at random from those available. Because automata move one at a time in a defined order, there are no "double kills" where one automaton may end up shooting at an enemy automaton that has already been killed.

For the model scenario discussed here, kills are determined by a simple system where, if a target is within range of an automaton's weapon, then it will be killed with a certain probability.

While the MANA model also allows somewhat sophisticated behaviour to occur by the use of "triggers" and more sophisticated layers of rules, the simple scenario explored in this report does not make use of these. Additionally, parameter settings for the scenario will only briefly be discussed. This is because from the point of view of the Metamodel, it is only important to characterise the positions of the automata as the models evolve, rather than concern ourselves with how the automata got to those positions (i.e. the rules of movement).

## 3 COMPARISON OF MANA DATA AND THE METAMODEL

### 3.1 The MANA cellular automaton model

While only one scenario is discussed in detail here, a number of different scenarios were explored for their consistency with the Metamodel. For the scenarios used, the entities within the model were given relatively low kill probabilities in order to allow enough time for self-organisation to occur before one or other force was eliminated. As noted above, if the single-shot kill probability is too high, one or other force may be eliminated very shortly after contact.

The value used was 0.01, meaning in a given time step in which a target automaton is within the shooting range of another automaton, there is a 0.01 chance of the target being killed in that time step.

Note that this probability does not necessarily correspond directly to the killing rate parameter, $k$, in Equation 1. There are other parameters that can affect the killing rate, such as firing range. For example, in the case where the two forces have uneven firing range, the force with the longer range will have a higher effective kill rate than its opponent even if both have the same killing probability. However, for the scenarios studied in this report, the firing range of both sides was always the same, and *k* for Equation 7 was simply taken to be the single shot kill probability.

This reports discusses the results from the "Meet" scenario previously introduced by Witty (Witty and Moffat, 2002). For other scenarios, it was typically found that the Metamodel equation worked at least as well as the Lanchester equations. It is hoped to describe further examples in a later report. Generally speaking, though, there is a large amount of variation possible in scenario set-up, making it impossible to explore all possibilities in this report.

Additionally, how well a scenario is described by the Metamodel will depend on how well it matches its assumptions. For example, scenarios where automata from one side fail to make a contribution to attrition (if, say, they remain in an area where contact with the enemy never occurs) but nonetheless count towards the fractal dimension estimation, then it may be expected that that scenario would be poorly represented by the Metamodel.

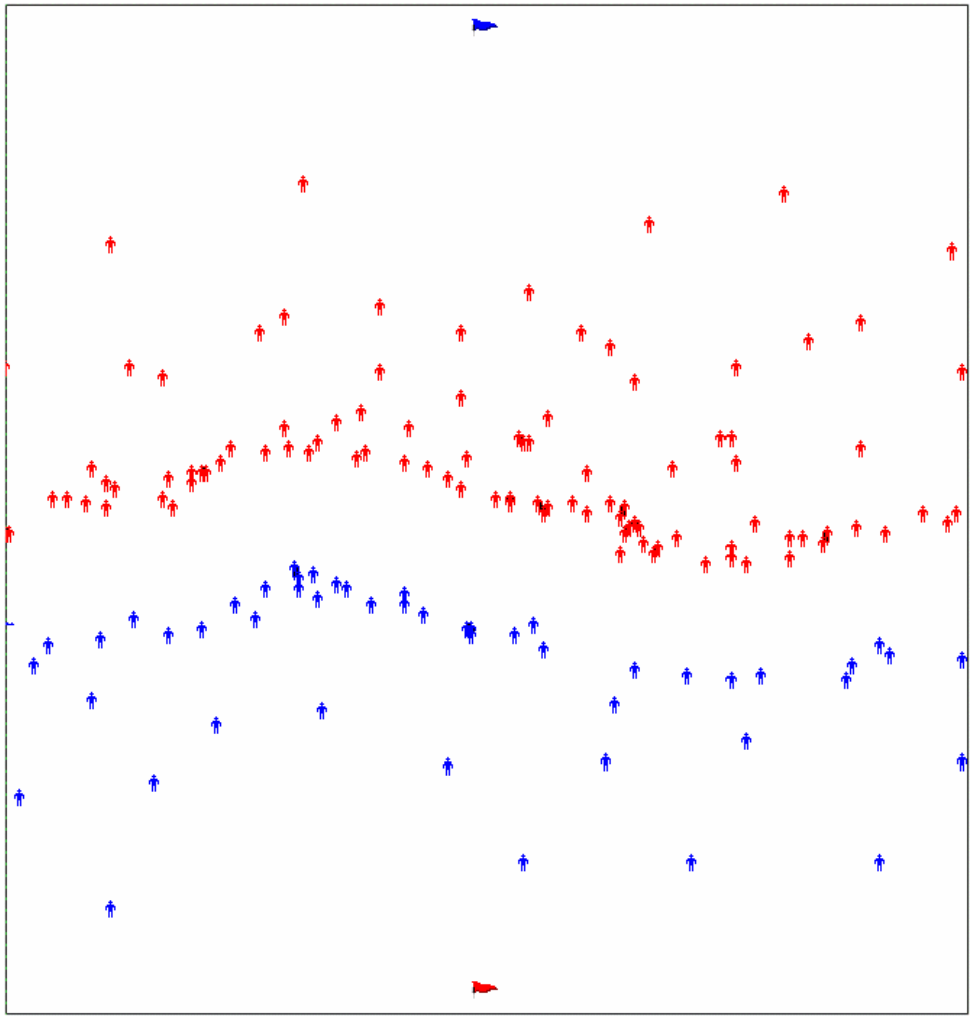

**Figure 3: Screen shot from the Meet scenario, showing the first phase of the run, where both sides form a "line", which displays a wave-like motion.**

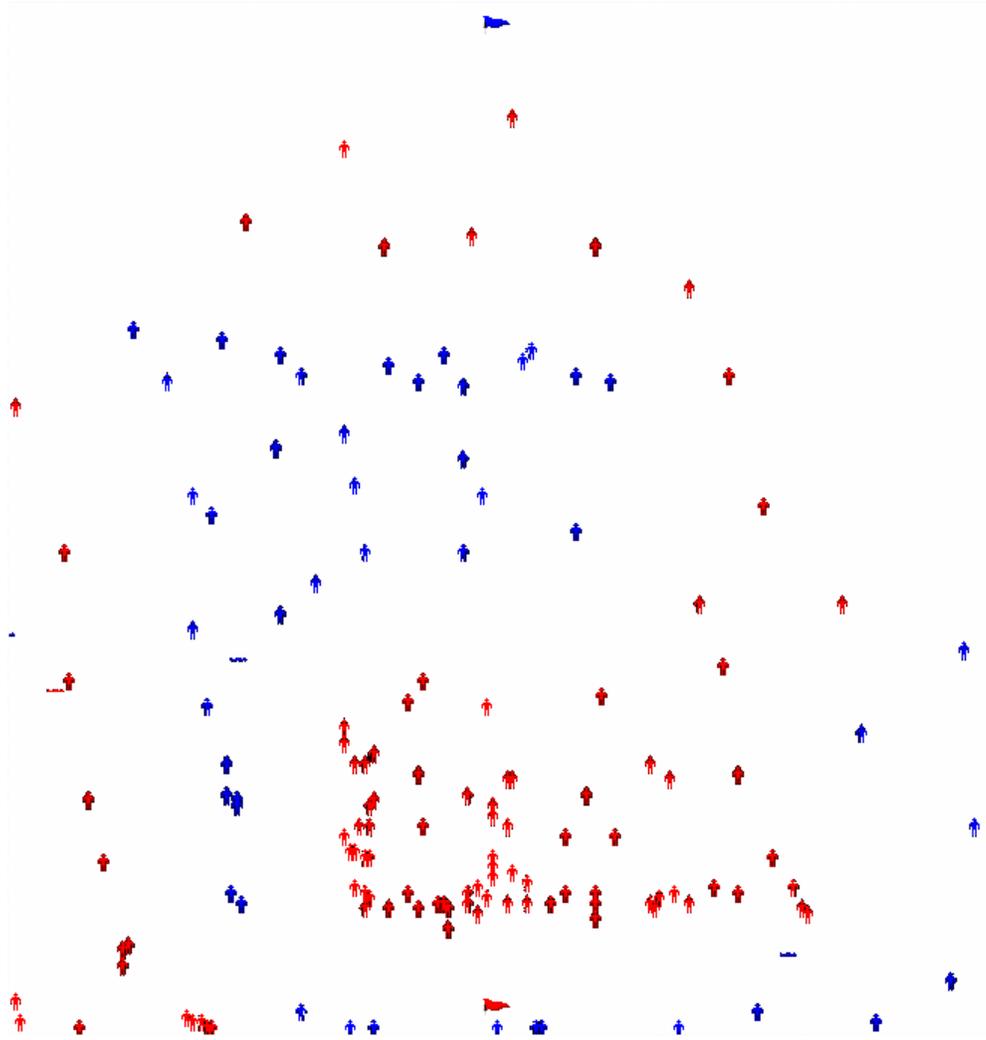

**Figure 4: Screen shot from the Meet scenario, showing the later phases of the same run shown in Figure 3.**

The Meet scenario is constructed so that the two sides meet in the centre of the battlespace and initially form a line, which moves in a wave-like manner. However, this line rapidly becomes unstable and breaks up into multiple clusters of automata after some unpredictable period of time. Figures 3 and 4 show the appearance of the distributions of Red and Blue automata at different stages of a single model run of the scenario.

The personalities of the automata are simple. Each automaton is repelled by other automata, and most strongly repelled by enemy automata (a weighting of –80 for enemy and –20 for friends). They are also attracted to move toward their waypoint (flag) at the opposite end of the screen (with a weighting of 10).

**3.2 Results**

Table 1 compares the predictions for LER from the Lanchester equation and the metamodel with the actual average outcome of the MANA model, using the Meet scenario. Here we look at different force ratios (i.e. the ratio of the starting numbers for Blue and Red) and variations on the value of $k$. The results are also compared with the Lanchester model, where this is given by:

$$\frac{dR}{dt} = -k_B B(t), \quad R(0) = R_0$$

$$\frac{dB}{dt} = -k_R R(t), \quad B(0) = B_0$$

Note that here we use the "square law" form of the Lanchester equation, which represents direct firing. Lanchester estimates were obtained by solving the above equations and estimating the final number of Red that would be expected to survive given that a certain level of Blue casualties had been reached.

| | $k_B$ | $k_R$ | $D_B$ | $D_R$ | $B_0$ | $R_0$ | $t$ | **MANA** | LER Lanchester | LER Metamodel | Size of factor of error Lanchester | Size of factor of error Metamodel |
|---|---|---|---|---|---|---|---|---|---|---|---|---|
| Case 1 | 0.01 | 0.01 | 0.82 | 1.05 | 50 | 100 | 200 | **1.1** | 2.3 | 0.54 | 2.1 | 2.0 |
| Case 2 | 0.01 | 0.01 | 0.81 | 1.23 | 30 | 100 | 200 | **1.0** | 5.3 | 0.35 | 5.3 | 2.9 |
| Case 3 | 0.01 | 0.01 | 1.14 | 1.23 | 80 | 100 | 200 | **1.0** | 1.3 | 0.83 | 1.3 | 1.2 |
| Case 4 | 0.03 | 0.01 | 1.22 | 1.38 | 50 | 100 | 150 | **0.42** | 0.68 | 0.26 | 1.6 | 1.6 |

**Table 1: Results from the MANA Meet scenario compared with the loss-exchange ratio expected using the Lanchester equation and the Metamodel.**

For the above cases, the Metamodel performed at least as well as the Lanchester estimates. This was typically also the case for the broader range of scenarios examined as part of this work, and for many of these the Metamodel performed significantly better than the Lanchester equation. However, this example was chosen here because it demonstrates how a numerically inferior force can be expected to improve its LER by adopting a more favourable distribution (and fractal dimension) than its opponent.

The results show that the LER is close to one for three of the cases, despite the numerical disparity between the two forces. This is not expected according to the Lanchester equation, but is explained by the Metamodel in terms of the fractal dimension for each force changing in such a way that Red causes a lower rate of attrition to Blue than Blue causes to Red. Significantly, even though the size of the factor of the error for the Metamodel in each case is not dissimilar to Lanchester, the Metamodel clearly suggests that Red's numerical advantage will not improve its LER as one might expect according to the Lanchester theory.

The investigation of other scenarios noted that, as a general rule, the Lanchester equation tended to work best when the automata maintained a single clump-like formation, or when there was a disparity in $k$ between the two sides.

# 4 CONCLUSIONS

It was shown how the equation proposed by the author could be used as a Metamodel for the expected LER for at least some cases of the MANA cellular automaton combat model. Generally speaking, this was particularly so for scenarios which displayed complex spatial patterns.

What was particularly interesting for the Meet scenario discussed in detail here was that the experimental MANA runs revealed that the LER maintained a value of around 1 for several different force ratios. It was also seen that the fractal dimension of each side changed as this force ratio changed, as would be expected according to the Metamodel if the LER was to remain constant.

While the Metamodel was capable of describing this phenomenon, the experimental results were at odds with the classical Lanchester theory. According to this, the LER ratio should change as the force ratio changes, because the potential firepower of one side has changed relative to the other. The Lanchester theory fails to describe the observed behaviour because it does not incorporate spatial distributions.

Thus it has been shown here that Equations 1 and 7 form a good Metamodel for the Meet scenario, while the Lanchester equation is poor. At the very least, the Metamodel appears to be useful for describing scenarios where at least one side has a complicated, dispersed spatial distribution. This may in turn allow investigation into the value of dispersed forces, guerilla and swarming tactics.

Further investigation is needed as to the realm of applicability of the model, and into how or whether it is appropriate to incorporate realistic effects like area fire into it.

# 5 ACKNOWLEDGEMENTS

The author would like to acknowledge the valuable assistance of Josephine M. Smith in reviewing the theoretical framework and helping to develop the expression for the Loss-Exchange Ratio. Her assistance with conducting the MANA experiments and analyzing the results discussed here is gratefully acknowledged. I would also like to thank Professor James Moffat for his contributions to this work, and Dr Gregory McIntosh for developing the software for automatically calculating the fractal dimension for the MANA experimental runs. Finally, I would like to acknowledge the contributions and discussions which took place at the Ninth Project Albert International Workshop within the TTCP JSA TP3 work group. The syndicate included Ms Smith, Dr Michael Ling, Dr Leung Chim, Mr Alex Ryan, Dr Kevin Ng, Mr René Séguin, Mr Lawton Clites and Dr Gary Horne.

# 6 REFERENCES

Gill, A. W. 2004. Improvement to the Movement Algorithm in the MANA Agent-Based Distillation, *Journal of Battlefield Technology*, Vol 7, No 2, 19-22.

Ilachinski, A. 2001. *Cellular Automata: A Discrete Universe*, World Scientific Publishing Company.

Lauren, M. K. 1999. Characterising the Difference Between Complex Adaptive and Conventional Combat Models, Defence Operational Technology Support Establishment, New Zealand, *DOTSE Report 169*, NR 1335.

Lauren, M. K. 2001. Fractal Methods Applied to Describe Cellular Automaton Combat Models, *Fractals*, Vol. 9, 177-185.

Lauren, M. K. 2002a. Firepower Concentration in Cellular Automaton Combat Models – An Alternative to Lanchester, *Journal of the Operational Research Society*, Vol. 53, 672-679.

Lauren, M. K. 2002b. A Fractal-Based Approach to Equations of Attrition, *Military Operations Research*, Vol. 7, No. 3, 17-30.

Lauren, M. K., and Stephen, R. T. 2002a, Fractals and Combat Modeling: Using MANA to Explore the Role of Entropy in Complexity Science, *Fractals*, Vol. 10, No. 4, 481-490.

Lauren, M. K., and Stephen, R. T. 2002b, Map-aware Non-uniform Automata (MANA) – A New Zealand Approach to Scenario Modelling, *Journal of Battlefield Technology*, Vol. 5, No. 1, 27-31.

Moffat, J. 2003. *Complexity Theory and Network Centric Warfare*. The US Command and Control Research Programme, Office of the Sec Defense, DoD, USA.

Moffat J., Passman M., 2004, Metamodels and Emergent Behaviour in Models of Conflict, invited contribution, *Simulation Modelling Practice and Theory*, Vol 12, pages 579-590.

Moffat J. and Witty S., 2003, Experimental Validation of Metamodels for Intelligent Agents in Conflict, special edition on Modelling and Simulation; *Information and Security*. Centre for Security Studies and Conflict Research, Zurich, Switzerland Vol. 12 No 1.

Moffat J. and Witty S. 2002. Phase Changes in Metamodelling using the Fractal Dimension. *Information and Security: An International Journal* (Special Issue on Agent Based Technologies), Vol. 8, No. 1, Pages 57-62.

Moffat J., Smith J., and Witty S., 2004, Emergent Behaviour; Theory and Experimentation Using the MANA Model, submitted to *Fractals*.

Witty S. and Moffat J. 2003, Initial Findings on Global Behaviour of an Agent Based Combat Model; An Investigation of the ISAAC Model, Published as a chapter of the book *Swarming : Network Enabled C4ISR*, Pages 13-14, sponsored by Assistant Secretary of Defense (Networks and Information Integration), US DoD.